\documentclass[a4paper,twocolumn,11pt,unpublished]{quantumarticle}
\pdfoutput = 1
\usepackage{graphicx}
\usepackage[colorlinks=true,citecolor=blue]{hyperref}
\usepackage[utf8]{inputenc}
\usepackage{enumerate}
\usepackage{algorithm2e}
\usepackage{amsmath}
\usepackage{multirow}
\usepackage{amssymb}
\usepackage{bbm}
\usepackage{color}
\usepackage{dcolumn}
\usepackage{bm}
\usepackage{amsbsy}
\usepackage{float}
\usepackage[normalem]{ulem}
\usepackage{braket}
\usepackage{changepage}
\usepackage[subrefformat=parens,labelformat=parens,caption=false]{subfig}
\usepackage{bbm}
\usepackage[dvipsnames]{xcolor}
\usepackage{esint}
\usepackage{lineno}
\usepackage{verbatim}
\usepackage{bm}
\usepackage{bbold}
\usepackage{placeins}
\usepackage{mwe}
\usepackage[numbers,sort&compress]{natbib}
\usepackage{amsmath,stmaryrd,graphicx}

\usepackage[subrefformat=parens,labelformat=parens,caption=false]{subfig}

\usepackage[export]{adjustbox}

\usepackage{soul}

\usepackage{tikz,xcolor}

\definecolor{lime}{HTML}{A6CE39}
\DeclareRobustCommand{\orcidicon}{%
	\begin{tikzpicture}
	\draw[lime, fill=lime] (0,0) 
	circle [radius=0.16] 
	node[white] {{\fontfamily{qag}\selectfont \tiny ID}};
	\draw[white, fill=white] (-0.0625,0.095) 
	circle [radius=0.007];
	\end{tikzpicture}
	\hspace{-2mm}
}

\foreach \x in {A, ..., Z}{%
	\expandafter\xdef\csname orcid\x\endcsname{\noexpand\href{https://orcid.org/\csname orcidauthor\x\endcsname}{\noexpand\orcidicon}}
}

\RestyleAlgo{ruled}

\DeclareMathOperator*{\argmin}{arg\!min}

\newcommand{\Rom}[1]{\uppercase\expandafter{\romannumeral #1\relax}}
\newcommand{\rom}[1]{\lowercase\expandafter{\romannumeral #1\relax}}

\begin{document}

\title{Improved Qubit Routing for QAOA Circuits}

\author{Ayse Kotil}
\affiliation{IQM, Georg-Brauchle-Ring 23-25, 80992 Munich, Germany}
\affiliation{Technical University of Munich, CIT, Department of Computer Science, Boltzmannstr. 3, 85748 Garching, Germany}
\author{Fedor \v{S}imkovic IV}
\affiliation{IQM, Georg-Brauchle-Ring 23-25, 80992 Munich, Germany}
\email{fedor.simkovic@meetiqm.com}
\author{Martin Leib}
\affiliation{IQM, Georg-Brauchle-Ring 23-25, 80992 Munich, Germany}
\email{martin.leib@meetiqm.com}

% \email{latex@quantum-journal.org}
% \orcid{0000-0003-0290-4698}
% \thanks{You can use the \texttt{\textbackslash{}email}, \texttt{\textbackslash{}homepage}, and \texttt{\textbackslash{}thanks} commands to add additional information for the preceding \texttt{\textbackslash{}author}. If applicable, this can also be used to indicate that a work has previously been published in conference proceedings.}

\maketitle

\begin{abstract}
% In order to run quantum circuits on superconducting quantum processing units (QPU) with limited connectivity, quantum information needs to be shuttled with the help of SWAP gates. Transpiling a circuit to be compliant with a given QPU topology necessitates the addition of SWAP gates and leads to increased circuit depth, both of which introduce additional errors in noisy quantum systems. Qubit routing has proven itself to be a hard task to solve optimally in terms of SWAP count and circuit depth due to the large search space for possible routing permutations. We develop a heuristic qubit routing algorithm for Quantum Approximate Optimization Algorithm (QAOA) circuits as well as an initial mapping method between logical and physical qubits. We empirically show that our algorithm is capable of producing quantum circuits with smaller SWAP gate counts and circuit depths compared to existing routing algorithms for QAOA circuits defined on $k$-regular or Erdös-Renyi graphs.

%In order to run arbitrary quantum circuits on quantum processing units (QPU) with limited connectivity, quantum information needs to be shuttled with the help of SWAP gates. Transpiling a circuit to be compliant with a given QPU topology necessitates the addition of SWAP gates which leads to increased circuit depth and presents a source of additional errors in non-error-corrected quantum processors. Finding near-optimal solutions to qubit routing is proven to be classically-hard, yet any improvement can potentially have a significant impact on the simulability of quantum algorithms of interest. 
We develop a qubit routing algorithm with polynomial classical run time for the Quantum Approximate Optimization Algorithm (QAOA). The algorithm follows a two step process. First, it obtains a near-optimal solution, based on Vizing's theorem for the edge coloring problem, consisting of subsets of the interaction gates that can be  executed in parallel on a fully parallelized all-to-all connected QPU. Second, it proceeds with greedy application of SWAP gates based on their net effect on the distance of remaining interaction gates on a specific hardware connectivity graph. Our algorithm strikes a balance between optimizing for both the circuit depth and total SWAP gate count. We show that it improves upon existing state-of-the-art routing algorithms for QAOA circuits defined on $k$-regular as well as Erdös-Renyi problem graphs of sizes up to $N \leq 400$.

\end{abstract}

\section{Introduction}

%premised on `quantum advantage', a term referring to the ability of quantum computers to solve some useful tasks that are intractable for classical computers. This advantage can manifest itself with exponential (as in Shor's Algorithm~\cite{shor1999polynomial}), quadratic (Grover's Algorithm~\cite{grover1996fast}) or super-polynomial computational speedup.

Quantum computing applications sparked recent interest in academia and industry, offering solutions for problems found in chemistry~\cite{mcardle2020quantum,elfving2020will}, finance~\cite{stamatopoulos2022towards} and optimization~\cite{blekos2023review}. One of the most promising applications of quantum computing is the speedup and improvement of the quality of solutions for combinatorial optimization problems. Combinatorial optimization use cases are ubiquitous in the industrial as well as academic research context and any quantum advantage would create a huge impact. However, the promise of a quantum advantage comes with a demanding engineering challenge: The presence of noise currently limits the number of operations one can perform on a quantum computer such that large-scale algorithms cannot be realized on Noisy Intermediate-Scale Quantum (NISQ) devices.
Therefore, QAOA is considered to be a promising candidate for near-term quantum computer applications due to shallow circuit depths.

However, the execution of QAOA circuits on many state-of-the-art quantum processors is hampered by either restricted local connectivity of the QPU or by the amount of two-qubit gates that can be executed in parallel. These constraints lead to the necessity of introducing SWAP gates to make the execution of interaction gates between non-neighboring qubits possible. This process is referred to as \textit{qubit routing} and designing a routine to map a quantum circuit to an architecture is often referred to as solving the qubit routing problem. Prior to the routing, the association of qubits in the abstract circuit description and actual, physical qubits on the QPU is referred to as \emph{qubit mapping} and there is, similar to the qubit routing problem, an exponentially growing number of possibilities to fulfill the mapping. The actual, implemented mapping, however, has a big influence on the overall algorithm performance, especially for NISQ devices. In order to bring the current state-of-the-art quantum computers closer to the practicability of large circuits, it is therefore essential that the qubit routing as well as the mapping is implemented in the most optimal way in terms of the number of added SWAP gates as well as resulting circuit depth. 

The qubit routing problem can be mapped to either the integer linear programming~\cite{Bhattacharjee2017} or the token swapping problem~\cite{Siraichi2019}, both of which are NP-complete~\cite{Miltzow2016, Karp1972}. However, an exponential classical run time of qubit routing would jeopardize the potential for quantum advantage within any quantum algorithm for which it is needed. Therefore, finding polynomial run time algorithms to achieve near-optimal results for qubit routing is the main focus of the present work.

Existing methods for qubit routing are diversified over different optimization objectives and solution approaches. Objectives can include minimizing the total number of gates of a certain type, minimizing the total circuit depth or reducing the overall error by taking noise models into account~\cite{Murali2019,Tannu2019,Ash-Saki2019,Siyuan2020, hashim2022optimized}. Learning-based methods have been explored in the form of deep learning~\cite{Acampora2021} and reinforcement learning~\cite{Pozzi2022}. The former approach is subject to data preparation overhead for the learning process and produces only architecture-specific results, while the latter suffers from long execution times and optimizes only for the circuit depth. Swap-network based qubit routing algorithms~\cite{hirata2011efficient,Kivlichan2018quantum, o2019generalized, Steiger2019, Weidenfeller2022} apply predefined layers of SWAP gates, achieving low circuit depths at the cost of high SWAP numbers. Swap-search based qubit routing algorithms~\cite{Zhu2020, Cowtan2019, Childs2019, Wille2019, Zulehner2017, Siraichi2018, Li2019, Gushu2019, Lingling2021} instead focus on selecting a subset of possible SWAP gates and evaluating the SWAP candidates according to a heuristic cost function, which generally leads to lower SWAP counts at the price of lower parallelism. The algorithm developed in this work follows this scheme as well, but attempts to strike a balance between the two aforementioned objectives.

In our work, we tackle the qubit mapping and routing problem for QAOA circuits. QAOA circuits exhibit a large search space to route interaction gates from, as all two-qubit gates commute with each other. This degree of freedom makes the routing task especially challenging, but also presents an opportunity to save gates and run time by a clever routine that picks favourable gates depending on the physical coupler layout of the QPU. We develop a heuristic algorithm which addresses both the circuit depth and SWAP number optimization and show that it outperforms currently used state-of-the-art routing algorithms, especially when simultaneously considering those two objectives.

The outline of this paper is as follows: In \autoref{sec:QAOA}, we briefly present the theoretical background of QAOA. \autoref{sec:method} introduces our routing algorithm followed by a discussion of numerical results presented in \autoref{sec:results}.

%QC is great 
%hardware is not commensurate with abstract circuit definition -> need for %routing
%routing is NP hard problem -> don't solve it exactly 

%Competition?
%How are we improving beyond the state of the art

%short description of the content of the article 
%we develop a general routing algorithm with known scaling 
%and derive from that a greedy heuristic that outperforms (in terms of total %numbers not scaling) it for small problem instances

\section{QAOA}\label{sec:QAOA}
The QAOA algorithm~\cite{Farhi2014} produces approximate solutions for combinatorial optimization problems. Combinatorial optimization problems pose the task of finding a binary number $x^*$ that minimizes a function $f: \{0,1\}^n \rightarrow \mathbb{R}$, $f_{min} = f(x^*)$. The way this problem is approached with QAOA is to first map the function $f$ onto a diagonal spin glass Hamiltonian  
\begin{equation}
H_P = \sum _{i<j} J_{ij}\sigma_i^z \sigma_j^z + \sum _{i} h_{i}\sigma_i^z\,,
\end{equation}
such that $f(x) = \bra{x} H_P \ket{x}$, where $\ket{x}$ is the computational basis state corresponding to bitstring $x$. Measurement samples from a low energy state of $H_P$ therefore are good solutions to the original combinatorial optimization problem. QAOA prepares this low-energy state by making use of the Rayleigh-Ritz variational principle, 
\begin{equation}
    f_{\text{min}} \leq \bra{\psi(\boldsymbol{\beta},\boldsymbol{\gamma})} H_P \ket{\psi(\boldsymbol{\beta},\boldsymbol{\gamma})}\,,
\end{equation}
where the parameters $\boldsymbol{\beta}=\{\beta_1, \dots, \beta_d\}$ and $\boldsymbol{\gamma}=\{\gamma_1, \dots, \gamma_d\}$ are optimized in an outer, classical, optimization loop. The QAOA Ansatz state is inspired by a trotterized version of quantum annealing with a simple, local driver Hamiltonian $H_0=-\sum_{i=1}^N \sigma_i^x$, where the length of the trotter steps are the variational parameters, 
\begin{multline}
       \ket{\psi(\beta,\gamma)} =  U(H_P, \beta_d)U(H_0, \gamma_d)\dots \\ \dots U(H_P, \beta_1)U(H_0, \gamma_1)\ket{+} \,,
\end{multline}
with $\ket{+} \propto (\ket{0}+\ket{1})^{\otimes n}$ the normalized, equal superposition of all computational basis states. 
The QAOA Ansatz circuit is composed of $d$ layers, each one consisting of a rotation generated by the problem Hamiltonian, $U(H_P, \beta_k)=e^{-i\beta_k H_P}$, followed by a rotation generated by the driver Hamiltonian, $U(H_0, \gamma_k)=e^{-i\gamma_k H_0}$. 

Since single qubit gates need no routing on any hardware platform the main target for routing is the parameterized rotation generated by the problem Hamiltonian. If we can find a routing for one of these unitaries we can always invert the routing for the next block in the Ansatz and iterate this way through the circuit. All 2-local terms in the problem Hamiltonian are products of two Pauli $z$ operators, which means that this many-body gate can be decomposed into a set of two-qubit interaction gates, $R_{zz} = \exp(-i \theta z_i z_j)$,  for any pair of qubits $i$ and $j$. Since all of these two-qubit gates commute, we can freely choose the order of their execution. The full information necessary for the routing process of the quantum circuit can thus be captured in a problem graph $\mathcal{G}_P=(V, E)$ with vertices $V=\{v_1,\dots v_N\}$ representing the qubits, and edges $E\subseteq\{e_{ij} = (v_i, v_j) \mid v_i, v_j \in V\}$ representing interaction gates with non-vanishing coupling strengths $J_{ij}$, where $J_{ij}$ is the weight associated with edge $e_{ij}$. If pairs of physical qubits, where the information for qubit $i$ and $j$ currently reside, are not connected with a physical coupler we have to shuttle quantum information with the help of SWAP gates using the techniques detailed in the following section.

% a class of problems where the solution is a bit-string satisfying clauses of logical boolean expressions subject to an objective function. An exhaustive search through all possible bit-strings results in an exponential run time, making the problem intractable for classical computers.

%In practice, one is interested in solution strings that do not satisfy all clauses but a subset of them, whose cardinality is to be maximized.

\section{Method} \label{sec:method}

\begin{figure}
 \center
  \includegraphics[width=8cm]{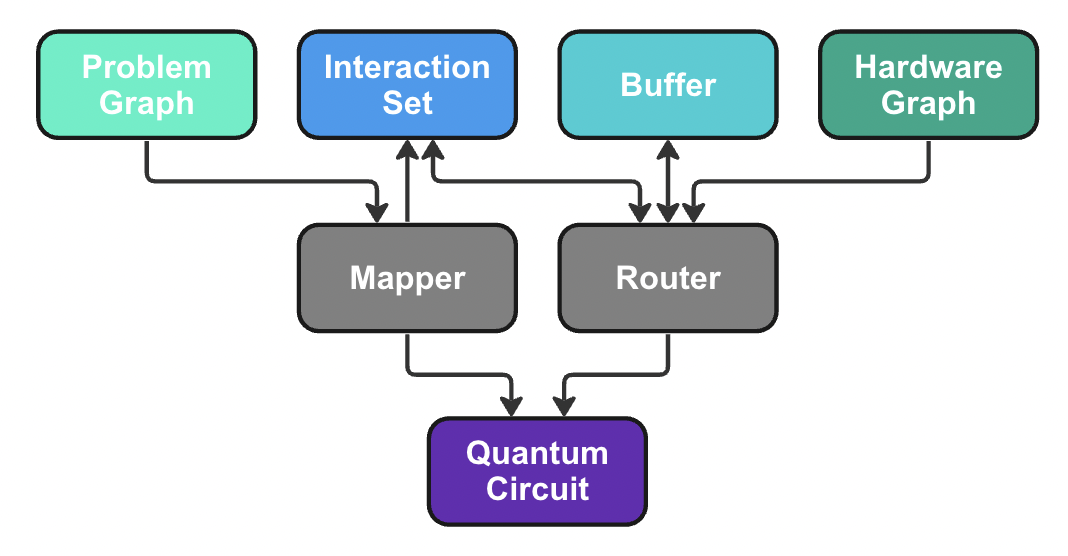}
  \caption{General setup for the routing algorithm introduced in Section \ref{sec:method}.}
  \label{fig:scheme}
\end{figure}

A \textit{quantum compiler} can take on various processing steps to execute a given quantum circuit on a QPU including, but not restricted to, decomposing circuit gates into native quantum gates, resynthesizing the circuit according to commutation and rewriting rules and adding SWAP gates to comply with hardware restrictions. In this work, we focus only on qubit mapping and routing. Our work can be integrated as a subroutine into an existing compiler responsible for other processing and execution steps. 

The general strategy of our algorithm is to first divide the set of interaction gates into a minimal amount of subsets such that every subset can be executed in parallel and without SWAP gates on a hypothetical all-to-all connected fully parallel QPU. Each of these subsets individually can, in principle, also be executed in parallel on a QPU with at least a one-dimensional topology and we take advantage of this fact in our mapping strategy. In order to execute the next of the remaining subsets we apply a greedy SWAP search routine to bring interacting pairs of qubits to sites on the QPU with a connecting coupler between them. 

Therefore, our QAOA routing algorithm consists of two main parts, the \emph{mapper} and the \emph{router}. An initial mapping is characterized by an injective function between the set of problem variables $x_i$ and physical qubits $q_i$. The problem variables will be referred to as \textit{virtual qubits}. The mapper's responsibility is to provide an initial mapping from virtual to physical qubits on a given QPU. We will consider only the settings where the number of physical qubits matches the number of virtual qubits for simplicity (bijective initial mapping). Having more physical than virtual qubits simply allows one to pick a given subset of them based on some given metric, like the connectivity or quality of the qubits. 

Once the qubits are initialized, the mapper will provide the router with an \emph{interaction set} of gates to route as well as an initial \emph{buffer} containing a set of gates that would be executable in parallel on an all-to-all connected QPU. The router will then attempt to execute the gates whilst adding SWAP gates in order to satisfy the connectivity constraints of the QPU, given by the \emph{hardware graph}. The interaction set of gates and buffer will be updated iteratively until all gates have been processed. The output of the compiler will be a quantum circuit logically equivalent to the input circuit from the problem graph and simultaneously compliant with a given QPU architecture. In the next subsections, we will discuss in detail how the mapper and router work by first introducing some preliminary definitions and then describing the procedures they follow. The full workflow of the algorithm is depicted in Fig.~\ref{fig:scheme}.

\subsection{Preliminaries}
In the following, we give formal definitions of some graph concepts which we use in designing our compiler.

An \textit{edge coloring} of a graph $\mathcal{G}$ is an assignment of edges to colors such that no adjacent edges are assigned the same color. The \emph{minimum edge coloring} is an edge-color mapping with the smallest possible number of colors. A \textit{matching} $M$ of a graph $\mathcal{G}$ is a subset of the graphs' edges where none of the edges are adjacent within the graph, i.e. the tuples containing pairwise non-adjacent edges. A \textit{maximum matching} $M_{max}$ is a matching with the maximum cardinality, i.e. a matching containing the maximum number of non-adjacent edges. A maximum matching is not necessarily unique. A \textit{maximal matching} is a matching that is not a subset of another matching, i.e. a maximal matching cannot accept another edge and still be a matching. Every maximum matching is a maximal matching, but not vice versa. A maximal matching can be obtained by traversing the edges of the graph randomly and adding them to a set if none of their vertices are already contained in the set.

Beyond the graph theoretical concepts, we will also need to define further objects used in the routing algorithm. In addition to the definition of the problem graph in section \ref{sec:QAOA}, every edge connecting virtual qubits stores the distance information between the physical qubits which the virtual qubits are assigned to prior to the routing. The distance information of an edge is updated if one of the virtual qubits is swapped. Finally we define $\mathcal{G}_{\text{QPU}}$ as the \emph{connectivity graph} of the QPU topology, where vertices are the physical qubits and an edge between two vertices exists if and only if there is a coupler between the corresponding qubits on the QPU (see also Fig.~\ref{fig:initial-mapping} for an example layout of the problem and connectivity graphs).
 We additionally maintain two containers for all interaction gates that need to be executed, i.e. the edges of the problem graph, $\mathcal{I}$ and $\mathcal{I}_C$. The \emph{buffer} $\mathcal{I}_C$ is a set of interaction gates where every involved qubit appears only once in an interaction gate. The target of the routing algorithm is to execute these gates and if not possible, because there is no suitable coupler in the QPU, enable them through the application of SWAP gates.  The rest of the interaction gates that have not been executed yet and don't belong to $\mathcal{I}_C$ are in the \emph{interaction set} $\mathcal{I}$. An execution of an interaction gate in the quantum circuit deletes it from $\mathcal{I}_C$ and triggers an exchange of gates between $\mathcal{I}_C$ and $\mathcal{I}$. The routing algorithm is finished when all interaction gates are erased from both $\mathcal{I}$ and $\mathcal{I}_C$. 

\subsection{Mapper}
\begin{figure}
    \centering
    \includegraphics[width=8cm]{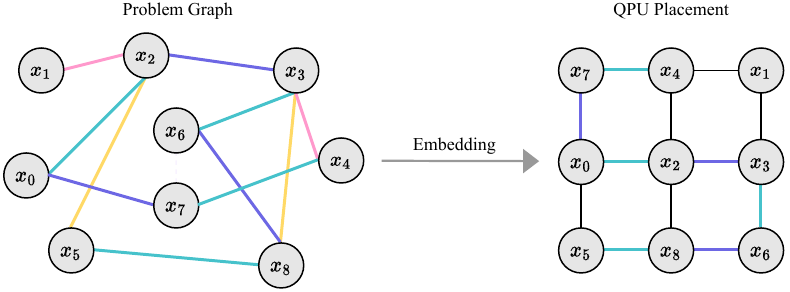}
    \caption{Edge coloring of a problem graph (\textit{left}) and the placement of virtual qubits on a physical square grid QPU (\textit{right}). The highlighted interaction gates can be executed within two layers without swapping.}
    \label{fig:initial-mapping}
\end{figure}

For an initial mapping, we determine two objectives: \rom{1}) After the initialization, every qubit is neighboring at least one other qubit it has to interact with and \rom{2}) The number of parallel interaction gates executable within the first two layers of the circuit is maximized. In order to achieve these objectives, we first determine the sets of concurrently executable interaction gates. This is equivalent to solving the edge coloring problem for $\mathcal{G}_P$. Finding an edge-coloring with the minimum number of required colors is an NP-hard problem. However, there exists a polynomial-time algorithm which guarantees a coloring with at most $\text{deg}(\mathcal{G}_P)+1$ colors, where $\text{deg}(\mathcal{G}_P)=\max({\text{deg}(v)| v\in V(\mathcal{G}_P)})$, which is at most one color worse compared to the optimal edge-coloring, according to Vizing's theorem~\cite{Vizing1965}.

 Once a coloring of $\mathcal{G}_P$ is found, we extract two colors and their respective edge lists. Next, we `chain' two colors together into a single list in an alternating manner, i.e. $C = \{(v_0,v_1), (v_1,v_2), \dots \mid (v_{2k}, v_{2k+1}) \in E_{{col}_0}, (v_{2k+1}, v_{2k+2}) \in E_{{col}_1})\}$. The chain will either be a cycle or a simple path, if necessary. In the case of a cycle, we can break it an arbitrary qubit pair to produce a simple path. In the case of multiple independent components of the induced subgraph resulting from these two colors, we can always connect them into a single chain at a low cost in terms of edges which must be placed back into $\mathcal{I}$. Next, the chain is embedded into the QPU hardware graph. For example, on a square grid QPU, the embedding starts with the bottom row of qubits and propagates to the top from right to left and vice versa, resulting in a continuous snake pattern. For QPU's where embedding of a chain is not possible (i.e. for tree graphs) because no Hamiltonian path exists, one must find an embedding by allowing for a minimal number of additional connections. Interaction gates corresponding to these connections then cannot be implemented and have to be placed back into the set of remaining interaction gates $\mathcal{I}$. 
 
 Fig.~\ref{fig:initial-mapping} shows an example of the initial mapping. An edge coloring of the problem graph (left) is performed and subsequently two colors are chosen (blue and green), which together form a chain that can be embedded into the square hardware connectivity graph (right). Consequently, all interaction gates from these two colors can be implemented concurrently in the first two circuit layers without the necessity for any SWAP operations. Assuming the QPU topology has a Hamiltonian path, this described allocation procedure allows for the execution of two sets of parallelizable interaction gates, one after the other, without any need for routing. By choosing the two colors with the largest edge sets, we can further increase parallelization.

\subsection{Router}\label{sec:router}
%describe edge coloring and token swapping idea 
%describe simplified greedy algorithm

% The routing algorithm takes as input two graphs: The problem graph $\mathcal{G}_P$ and the QPU connectivity graph $\mathcal{G}_{\text{QPU}}$. For $\mathcal{G}_P$, the problem variables are mapped to vertices and variable interaction gates are mapped to edges: $V_{\mathcal{G}_P}=\{v_i \equiv x_i \mid i \in \{1,\dots, N\}\}$, $E_{\mathcal{G}_P}=\{(v_i, v_j) \mid x_i, x_j \in P \}$. For $\mathcal{G}_{\text{QPU}}$, the qubits are mapped to vertices and the QPU coupling is represented by edges: $V_{\mathcal{G}_{\text{QPU}}}=\{v_i \equiv q_i \mid i \in \{1,\dots, N\}\}$, $E_{\mathcal{G}_{\text{QPU}}}=\{(v_i, v_j) \mid \text{$q_i,  q_j$ are connected}\}$. We calculate the shortest paths between qubits of $\mathcal{G}_{\text{QPU}}$ prior to routing. This needs to be done only once (e.g. with Dijkstra's algorithm). 

The routing algorithm takes as input: $\mathcal{I}$, $\mathcal{I}_C$, the initial mapping of virtual to physical qubits and the QPU connectivity graph $\mathcal{G}_{\text{QPU}}$. We calculate the shortest paths between physical qubits of $\mathcal{G}_{\text{QPU}}$ prior to routing. This needs to be done only once (e.g. with Dijkstra's algorithm). The router then maintains the buffer $\mathcal{I}_C$ and will gradually proceed with adding the remaining qubit interaction gates from $\mathcal{I}$ into this set according to update rules to be described later on. The routing algorithm consists of the following main steps: 
\begin{enumerate}[I]
  \item Execute all possible interaction gates which have a valid connectivity on the QPU and remove them from $\mathcal{I}_C$.
  \item Trigger the update and refilling strategy between $\mathcal{I}$ and $\mathcal{I}_C$
  \item For all the interaction gates which do not have a valid connectivity on the QPU, apply the swap strategy.
  \item Repeat \Rom{1}-\Rom{3} until both $\mathcal{I}$ and $\mathcal{I}_C$ are empty. 
\end{enumerate}

\begin{figure}[t]
  \centering
  \includegraphics[width=8cm]{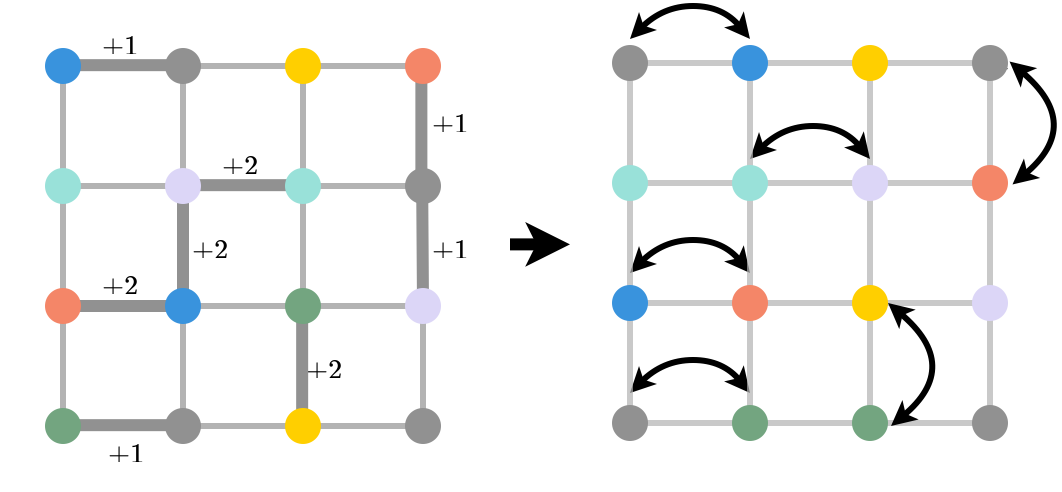}
  \caption{Given the allocation of the qubits on the QPU and the interaction gates, indicated by matching colors, the maximum matching of the resulting SWAP graph leads to the allocation of the physical qubits as shown on the right side. The qubits not included in $\mathcal{I}_C$ are indicated by gray coloring.}
  \label{fig:swap-graph}
\end{figure}

% \subsection{Dynamic Strategies}
%  We apply some update and removal rules, described below, to add remaining interaction gates as gates are removed from $\mathcal{I}_C$. That way $\mathcal{I}_C$ becomes a dynamic set that changes with each iteration loop and after each finished interaction gates.

\subsubsection{Swap Strategy}

 We define the total distance $\mathcal{D}$ of $\mathcal{I}_C$ to be the sum of the shortest path lengths between virtual qubit pairs of the interaction gates that are contained in the buffer $\mathcal{I}_C$. When two qubits are swapped on a QPU, it holds for the new distance $\mathcal{D}'$ that $|\mathcal{D}'-\mathcal{D}|\leq 2$. There are 5 possible outcomes:

\textbf{\rom{1} \& \rom{2}.}  Both of the swapped qubits are in $\mathcal{I}_C$ and they move closer to (further apart from) their partners, i.e. $\mathcal{D}'-\mathcal{D}=2$ ($\mathcal{D}'-\mathcal{D}=-2$).

\textbf{\rom{3}.} Both of the swapped qubits are in $\mathcal{I}_C$ and one of them moves closer to its partner, whilst the other one moves further away i.e. $\mathcal{D}'-\mathcal{D}=0$. Alternatively, one can have $\mathcal{D}'-\mathcal{D}=0$ if none of the swapped qubits are in $\mathcal{I}_C$, which is a case not considered as a swap candidate. 

\textbf{\rom{4} \& \rom{5}.} Only one of the qubits is in $\mathcal{I}_C$ and it moves closer to (further apart from) its partner, i.e. $\mathcal{D}'-\mathcal{D}=1$ ($\mathcal{D}'-\mathcal{D}=-1$).

Ideally, the routing algorithm should only execute swaps that result in a decrease of the total distance which we will refer to as \textit{positive swaps}. Finding a maximum matching from all possible positive swaps has the objective of maximizing the amount of swaps that can be executed in parallel in order to reduce the overall circuit depth. Hence, the swap strategy consists of the following steps:

\begin{enumerate}[I]
  \item Find swap candidates by traversing through the edges of $\mathcal{G}_{\text{QPU}}$. 
  \item Construct a \textit{swap graph} $\mathcal{G}_{\text{SWAP}}$ consisting only of positive swaps.
  \item Find a maximal matching of $\mathcal{G}_{\text{SWAP}}$.
  \item Apply the swaps found by the maximal matching.  
\end{enumerate}

An example for a $4\times 4$ square QPU is shown in Fig.~\ref{fig:swap-graph}. Here, qubit pairs are indicated through matching colors and all swap candidates with a positive swap score are highlighted by thick edges. On the right side of Fig.~\ref{fig:swap-graph} the result of the maximal matching is presented, i.e. arrows indicate all swaps which have been be applied and the qubit colors show the resulting configuration. 

One needs to additionally pay attention to a number of corner cases involving pairs of swap candidates which cancel each others positive or negative effects. This can happen if they both change the distance of the same qubit pair. For simplicity, let us consider a square QPU connectivity graph. Now, if the horizontal or vertical distance between a pair of qubits is exactly one then separately applying swaps acting on both qubits in that given direction will have no net effect on their distance. To avoid this issue we apply swaps from the maximal matching one by one and perform an additional check of whether a given swap is still positive before implementing it. 
Another case concerns pairs of swap candidates which act on a qubit pair with a horizontal or vertical distance of zero (meaning they are located in the same row or column of the square QPU graph). Here, it can happen that, individually, both swap candidates in the given direction have a score smaller than $+1$, but if applied together their score becomes positive because the negative effect on the qubit pair cancels out. To this end, it is possible to additionally check for such swap candidate pairs (with swap scores $0/0$ or $-1,0$) after having applied all the positive swaps from the minimal matching. If all four qubits involved in the swaps have not been utilized in a given swap layer, one then implements the pair of swap candidates. After the swapping, the compiler continues to the next step and applies interaction gates that have now become executable on the given hardware. 

For the case that no candidates have been found to reduce the overall distance, but $\mathcal{I}_c$ is not empty, the fallback strategy is to pick a random qubit pair to swap, where one of the qubits is guaranteed to move closer to its partner after the swap (resulting in a $+0$ change in $\mathcal{D}$). 

% To see why this is necessary, consider the one-dimensional sequence of qubits $q_1, q_2, q_3, \bar{q}_1, \bar{q}_2$\, where qubits with the same index correspond to interaction gate (here $q_3$ is not part). It is clear that for this configuration all three possible swap candidates have a $+0$ score even though at least one of the swaps needs to be implemented to resolve the interaction gates.

\begin{figure*}
    \centering
    \includegraphics[width=15.5cm, height=8.5cm]{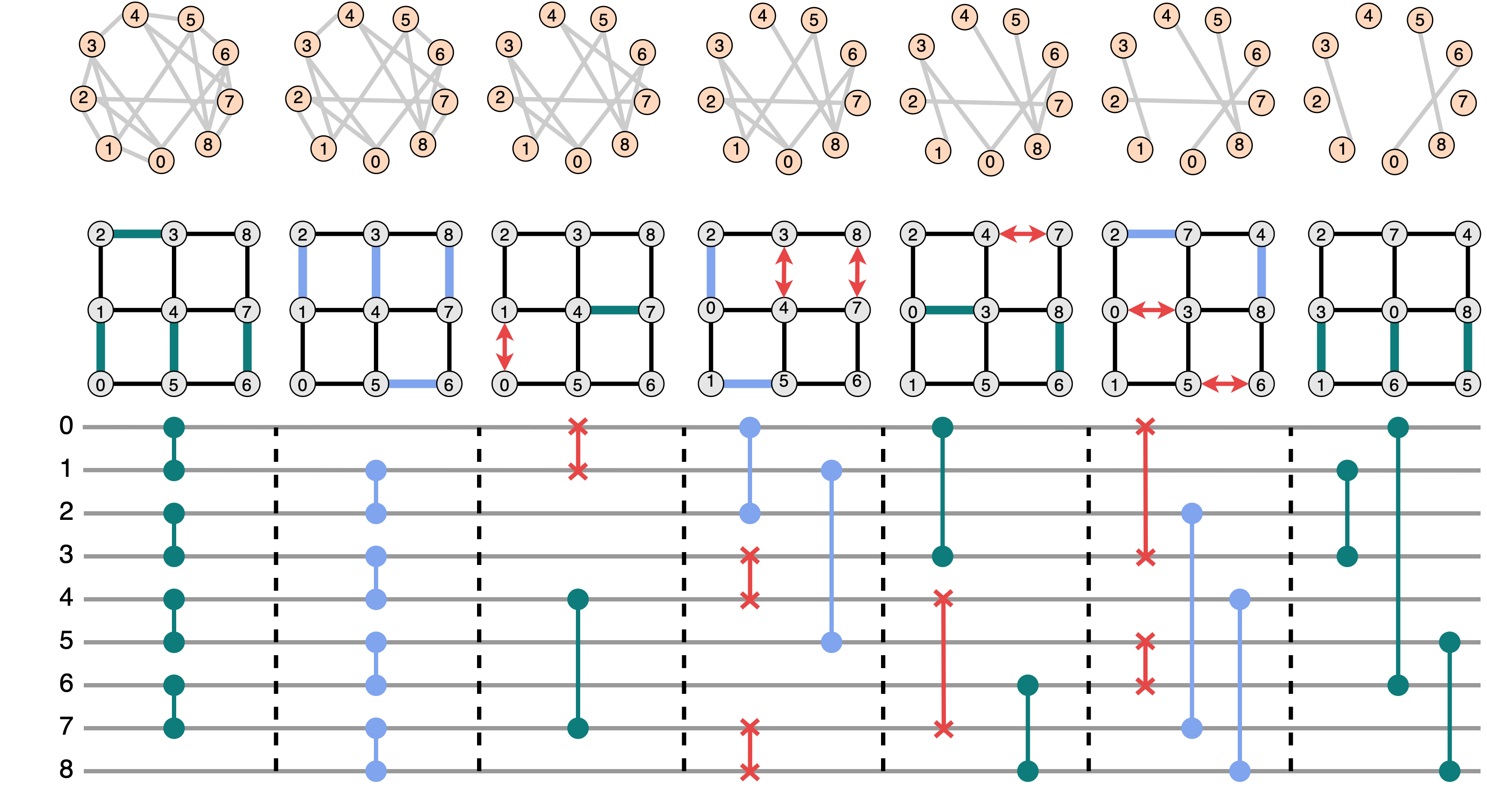}
    \caption{A 9 qubit example of the complete routing process of a 4-regular QAOA graph, showing the remaining interaction gates in the interaction set $\mathcal{I}$ (\textit{top}), the $3\!\times\!3$ square-grid QPU allocation (\textit{center}) and the quantum circuit (\textit{bottom}). The executed interaction gates are removed, step by step, from $\mathcal{I}$. Red gates correspond to SWAP operations added by the router, whereas blue and green gates correspond to $R_{ZZ}$ interaction gates. Dashed vertical lines indicate individual circuit layers. After seven circuit layers, all gates have been successfully executed.}
    \label{fig:big-pic}
\end{figure*}

\subsubsection{Update Strategy}
The update strategy consists of steps that aim to replace interaction gates in the buffer $\mathcal{I}_C$. As the routing proceeds, it may be of advantage to exchange some pairs in the current set $\mathcal{I}_C$ that are still far away from each other. The replacement pair should have a smaller distance than some other pair in  $\mathcal{I}_C$. In order to achieve this, we iterate through every qubit that is involved in an interaction gate from $\mathcal{I}_C$ and check the available neighbors of its respective vertex which are currently not involved in another gate within the buffer, $u$, in $\mathcal{G}_{P}$. We choose the neighbor with the minimal distance $W$ on the hardware graph $\mathcal{G}_{\text{QPU}}$, i.e. the partner with the smallest distance to the vertex. If 
\begin{equation}
\tilde{v} = \argmin_{v^\prime\in [N],v^\prime \neq v} W(v^\prime,u) 
\end{equation} 
then $(u, v)$ is replaced by $(u, \tilde{v})$ in the buffer $\mathcal{I}_C$.

\subsubsection{Removal Strategy}
Every removal triggers additions from $\mathcal{I}$ to $\mathcal{I}_C$, if applicable. When an interaction gate is routed successfully, its qubits become available for a further pair selection. We search through the interaction pairs involving the two qubits and where other logical qubits are not currently involved in another gate contained in the buffer $I_C$. From there, we pick the ones with the minimum current distance and add them to $\mathcal{I}_C$. 

\subsubsection{Push-back Strategy}

 In order to reduce the circuit depth, we check for every applied gate (including swap gates) whether it can be already executed in a previous layer. We accomplish this by maintaining a set of layers for every circuit for easy access. A layer is characterized by a graph where nodes correspond to physical qubits, and an edge exists if and only if two qubits are subject to an interaction gate within that layer. In addition, each gate will virtually `block' a newly added gate from being pushed back any further, which is indicated by a binary node attribute. The edges are labeled by either a swap or a non-swap mark. If two swap gates are inserted on the same pair of qubits sequentially, they will cancel each other out by a swap mark check. 
 If the qubits involved in a certain gate are both `free' in the previous layer, which can be checked by accessing the previous layer and the node attributes of the qubits, we push back the gate. This process is repeated until at least one of the involved qubits is involved in another gate in the previous layer.
 
 \subsection{Example}
 
 In Fig.~\ref{fig:big-pic} we show a full instance of the routing algorithm for a 9 qubit 4-regular connectivity graph embedded in a $3\times 3$ square lattice. We observe that the first two circuit layers, as described, are implemented without any swap operations. The remaining interaction gates (blue and green) are implemented in the following five layers using 6 SWAP gates (red). Note that SWAP and interaction gates can also appear in the same circuit layer as a result of the push-back strategy. The full circuit can be executed in depth 7.

\section{Numerical Results}\label{sec:results}

\begin{figure*}[t]
  \centering
  \includegraphics[scale=0.5]{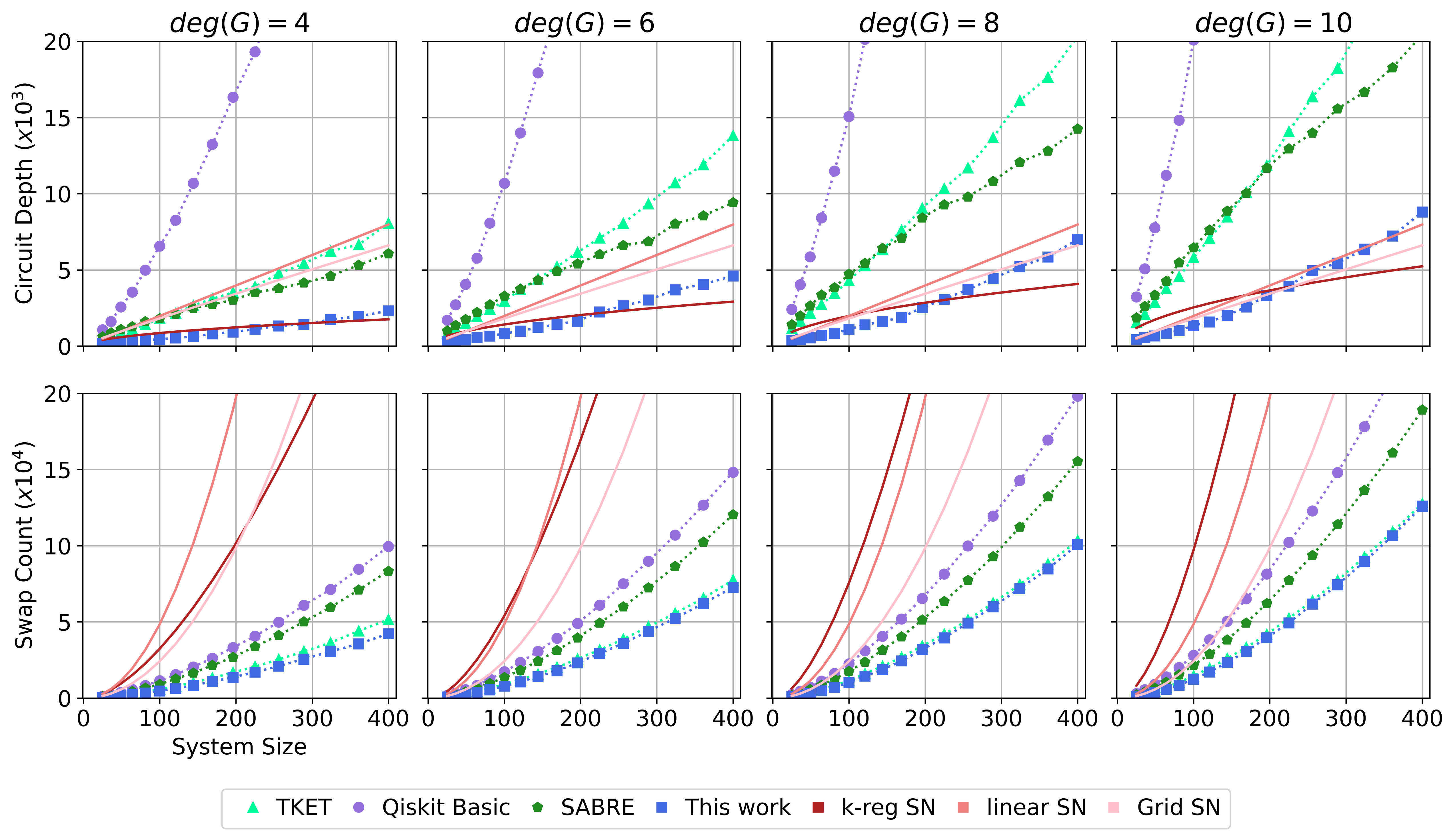}
  \caption{Comparison of circuit depths (top) and SWAP counts (bottom) for different routing method. Counts correspond to a single QAOA layer for $k$-regular graphs with $k\in \{4, 6, 8, 10\}$ and as a function of the number of qubits $N$ in the problem graph. Dotted lines correspond to numerical results whilst full lines are theoretical bounds for different SWAP networks.}
  \label{fig:gridplot-reg}
\end{figure*}

\begin{figure*}[t]
  \centering
  \includegraphics[scale=0.5]{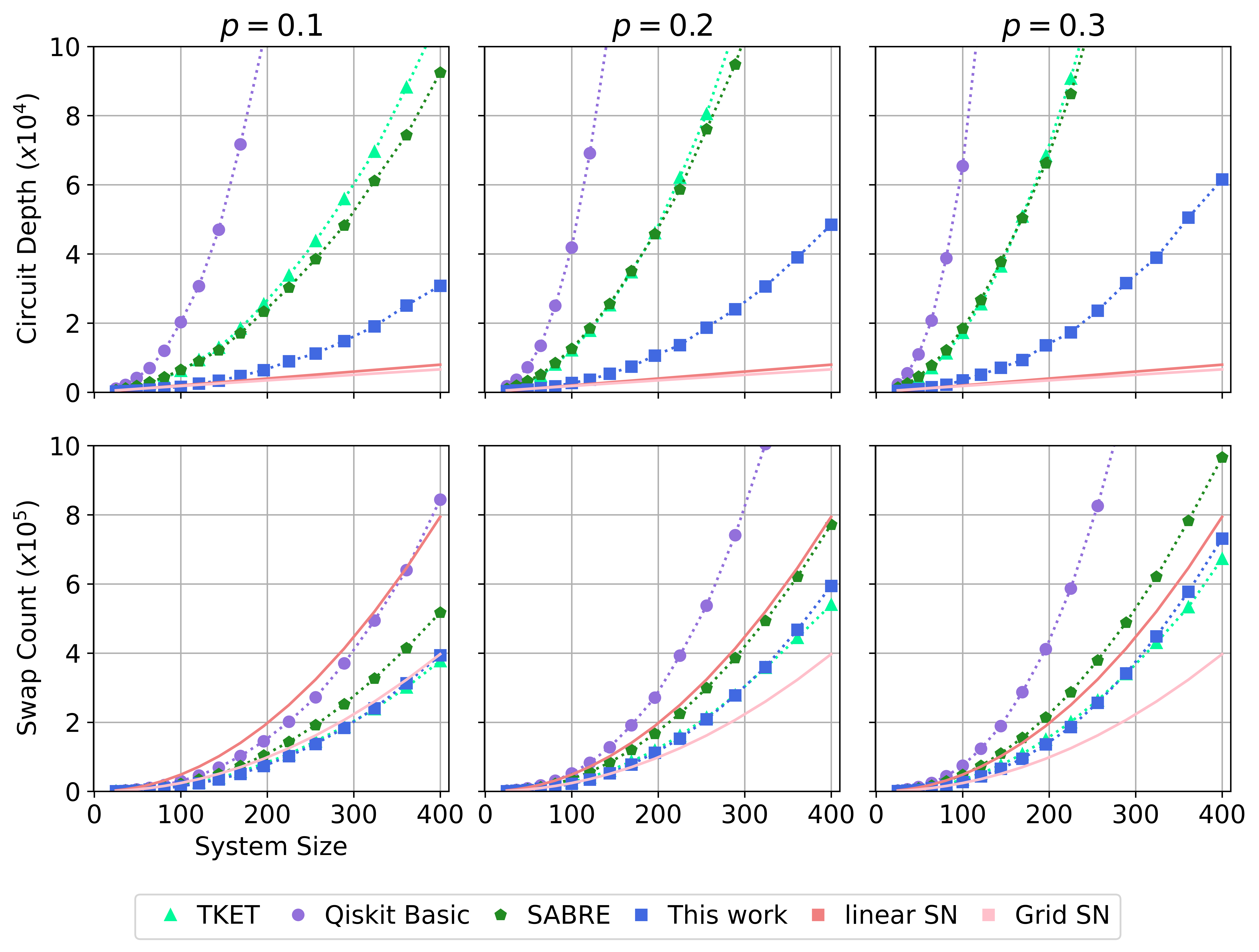}
  \caption{Comparison of circuit depths (top) and SWAP counts (bottom) for different routing method. Counts correspond to a single QAOA layer for random Erdős–Rényi graphs with gate probabilities $p\in \{0.1, 0.2, 0.3\}$ and as a function of the number of qubits in the graph. Dotted lines correspond to numerical results whilst full lines are theoretical bounds for different SWAP networks.}
  \label{fig:gridplot-random}
\end{figure*}

\subsection{Setup}

In this section we aim at benchmarking our routing method against other available algorithms found in literature. In what follows, we consider all single-qubit gates to be a free resource and any two-qubit gate to be implementable within a single circuit layer. Any two gates can be parallelly implemented within the same layer only if they are acting on distinct qubits.

We have conducted our comparison by considering randomized $k$-regular and Erdős–Rényi graphs as problem graphs. A graph is $k$-regular if every vertex has degree $k$. Erdős–Rényi graphs are constructed by adding edges randomly with a probability $p$. We have considered the routing algorithm for 4-, 6-, 8- and 10-regular graphs as well as for Erdős–Rényi graphs with probabilities $p = \{0.1, 0.2, 0.3\}$ and run all numerical experiments on a two-dimensional square-grid QPU with linear system size $L$ ranging from $5$ to $20$, i.e. the total number of qubits is $ N \in \{L^2 \mid L \in \{5,\dots,20\}\}$. We would like to emphasize that the choice of a square lattice hardware graph is arbitrary and our routing method is not limited to or optimized for this particular hardware graph. On the other hand, a square grid QPU corresponds to existing hardware realizations and the results are therefore likely most informative of the realistically attainable performance of the routing methods. In order to acquire sufficient statistics, we average over 20 random instances for each problem size (and graph type) whilst keeping the random seeds unchanged between different routing algorithms.

\subsection{Benchmark Algorithms}

%~\cite{Qiskit}
%~\cite{Cirq2022}
%~\cite{sivarajah2020t}

We have compared the circuit depth and the number of added SWAP gates for the following routing algorithms: Tket  (version: 1.10.0), Qiskit's implementation of SABRE~\cite{Li2019} and its own Basic routing algorithm (version: 0.39.4). We included only benchmarks from routing algorithms we could run in a reasonable amount of time for all system sizes up to 400 qubits, up to the order of hours on a moderately sized computing cluster. We did not include results from the algorithm provided by Cirq (version: 1.1.0) or the implementation of 2QAN~\cite{Lao2022} as they had execution times far above our stated threshold already for smaller system sizes. 

Additionally, we considered three SWAP network (SN) implementations, which are mostly agnostic to the problem graphs and simply shuffle around qubits within a given hardware graph in order to ensure that every pair of qubits is neighboring at a given time step. The first one is a linear even-odd SWAP network~\cite{hirata2011efficient}, which executes in $2N-2$ depth (including interaction gates) whilst implementing $N^2/2-3N/2+1$ SWAP gates. This SWAP network can be applied as long as a Hamiltonian path through the hardware graph exists. Another SWAP network, tailored to $k$-regular problem graphs and square grid hardware graph, has been introduced in Ref.~\cite{Steiger2019}. The algorithm is based on the fact that one can shuffle qubits to arbitrary positions on a square grid in order to implement a full layer of interaction gates in $3L$ circuit layers, which for a $k$-regular problem graph totals $3(k-1)N^{1/2}-2k+4$ depth and $3(k-1)(N^{3/2}/2-3N/2 +N^{1/2})$ SWAP operations. Finally, a SWAP network for the square grid was implemented in Ref.~\cite{Weidenfeller2022}.The authors have also defined SWAP network strategies for a few different hardware architectures: linear and heavy-hex. The square-grid version of the algorithm takes $3N/2+3N^{1/2}+1.5$ depth and $(N/2+N^{1/2}+1/2)(N/2-N^{1/2})$ SWAP operations. Note that, in general, two interaction gate layers are needed after every SWAP layer in this SWAP network.

\subsection{k-regular Graphs}

In Fig.~\ref{fig:gridplot-reg} we compare the performance of the aforementioned routing algorithms for $k$-regular graphs for $k=\{4,6,8,10\}$ and for system sizes of up to 400 qubits placed on square hardware graphs. We observe that our routing method produces significantly shallower circuits compared to all non-SN methods and the difference grows with both the system size and the regularity of the graphs (see top row of Fig.~\ref{fig:gridplot-reg}). On the other hand, we see that all SN's eventually produce favourable circuit depths compared to our method as the regularity and system size are increased. This is expected, since the linear and grid SN are unaffected by the number of interaction gates that need to be implemented and the $k$-regular SN only scales with $k$ and the square root of the total number of qubits, $\sqrt{N}$. Indeed, for small regularities ($\leq 10$) the $k$-regular SN is optimal at large enough system sizes ($\gtrsim 250$), whilst our method performs best for system size below. The trade-off of SN's becomes clear when the focus shifts to SWAP counts (bottom row), as they produce far higher values compared to non-SN algorithms. Our method has the lowest SWAP counts for all regularities, with only TKET performing comparably well, especially at higher regularities, and at the cost of much higher circuit depths. 

\subsection{Erdős–Rényi Graphs}
Let us now consider Erdős–Rényi graphs with interaction gate probabilities of $p = \{0.1, 0.2, 0.3\}$ in Fig.~\ref{fig:gridplot-random}. For these graphs, the number of interaction gates grows with the number of qubits and especially for large system sizes the graphs become significantly denser than those of Fig.~\ref{fig:gridplot-reg} (i.e. for 400 qubits the number of interaction gates with $p=0.1$ is comparable to a $40$-regular graph). Furthermore, for such graphs the k-reg SN is not appropriate, as it will scale with the node which has highest regularity, which can be as high as the system size  $N$. In general, we observe for these graphs that the SN's perform significantly better in terms of depth and also for the SWAP counts for $p=\{0.2,0.3\}$ whilst for $p=0.1$ our method is still optimal. Indeed, $p=0.1$ seems to be close to the break even point, up to where our method performs better than SN's. Of the two SN's the grid SN performs better for all of the considered graphs, however we expect the linear SN to take over at some point as $p\to 1$, since it is provably optimal in this limit. 

% \subsection{Conclusions}
\section{Discussion}\label{sec:discussion}

We conclude that our routing algorithm is especially useful for sparse problem graphs and for system sizes up to hundreds of qubits. Given the current technological progress in quantum hardware we can assume that exactly these kinds of QAOA instances bear the most promise to be successfully implemented in the near-future, possibly without the need for fault-tolerance. Furthermore, our algorithm is not restricted by the regularity of problem graphs or the connectivity of the hardware graphs. This means that one would be able to significantly improve the performance of our algorithm compared to SWAP network approaches by implementing QPUs with additional hardware connectivity.

The run time of our algorithm is limited by the polynomially scaling evaluation of the maximal matching routine~\cite{bernshteyn2023fast}. In principle, it is also possible to improve upon our algorithm by implementing ever more sophisticated qubit scores. For example, one could think of treating \emph{false} swap candidate pairs that have a mutually cancelling effect on the total swap distance $\mathcal{D}$ on the same footing as single swap candidates (currently the maximal matching is performed only over those). Alternatively, one could consider giving a positive contribution to the score of swaps if their unpaired qubits would move in the direction of the closest pair which is not in $\mathcal{I}_C$. Yet another option is to  prioritize pairs which are closer to each other by introducing a slight correction to the swap score, since further apart pairs are more likely to be replaced in the buffer $\mathcal{I}_C$. One could also further reduce the SWAP counts at the expense of increasing the circuit depth by only accepting $+2$ swap candidates as long as such candidates are available and only then resorting to $+1$ candidates. It is important to note at this point, however, that since our algorithm is heuristic in nature, it is possible to arbitrarily keep improving its swap and depth performance at the cost of increased complexity and execution times until one reaches the optimum at the price of an exponentially scaling algorithm. We believe that, in this sense, further improving our routing algorithm will likely only yield diminishing returns.

 We have introduced an efficient qubit routing strategy for QAOA which strikes a balance in simultaneously performing well in terms of swap counts as well as total circuit depths. We find the most striking improvement compared to existing methods for relatively sparse graphs of the order of a few hundred qubits. Beyond that our algorithms stays competitive in the number of swap gates, but succumbs to swap network algorithms in the metric of total circuit depth. Unlike some alternatives, our algorithm is agnostic to the actual structure of the hardware graph and simply improves with its connectivity. On the other hand, some of the swap networks need a specific connectivity, like a chain or square grid to be implemented and do not necessarily improve when additional connectivity is available. 

In this work, we considered every two-qubit operation to be directly implementable. In reality, however, SWAP gates are often not directly implementable on a QPU, in which case they have to be decomposed i.e. into a circuit involving three successive CNOT gates. Whilst this does not alter the relative efficiency we found for different routing algorithms, it adds additional emphasis on the importance of optimizing this part of QAOA.

In Ref.~\cite{itoko2020optimization} the authors considered bridge gates, which effectively implement a CNOT gate between next-nearest neighbors in depth four as an alternative to swapping and implementing the CNOT gate afterwards, which also results in a depth four circuit. We want to point out that this strategy does not work in the case of QAOA, since the necessary interaction gate here is $R_{zz}$. 

In some QPU architectures it is possible to shuffle around qubits without having to implement any actual swap gates, rendering the qubit connectivity effectively all-to-all. However, even then, the shuffling process can take significant time in which idling qubits can potentially suffer from noise. One can then view the qubit routing as a useful means of decreasing this idling times of qubits as well as the total run time of algorithms.

Whilst we have designed our method with the QAOA algorithm in mind, it can also be applied as a general circuit optimization strategy. Here, it is most powerful when given a circuit with mostly commuting gates, as one then has the choice of selecting the most efficiently implementable interaction gates for the buffer $\mathcal{I}_C$. For not commuting gates, on the other hand, one is obliged to keep the existing order of gates. circuit optimization, using related concepts has been previously explored in Ref.~\cite{itoko2020optimization}.

% https://arxiv.org/pdf/2111.04572.pdf

\section*{Acknowledgments} The authors would like to thank Stephanie Cheylan for useful discussions.

% \printbibliography

% \bibliographystyle{quantum}
\bibliography{main}

\begin{thebibliography}{34}%
\makeatletter
\providecommand \@ifxundefined [1]{%
 \@ifx{#1\undefined}
}%
\providecommand \@ifnum [1]{%
 \ifnum #1\expandafter \@firstoftwo
 \else \expandafter \@secondoftwo
 \fi
}%
\providecommand \@ifx [1]{%
 \ifx #1\expandafter \@firstoftwo
 \else \expandafter \@secondoftwo
 \fi
}%
\providecommand \natexlab [1]{#1}%
\providecommand \enquote  [1]{``#1''}%
\providecommand \bibnamefont  [1]{#1}%
\providecommand \bibfnamefont [1]{#1}%
\providecommand \citenamefont [1]{#1}%
\providecommand \href@noop [0]{\@secondoftwo}%
\providecommand \href [0]{\begingroup \@sanitize@url \@href}%
\providecommand \@href[1]{\@@startlink{#1}\@@href}%
\providecommand \@@href[1]{\endgroup#1\@@endlink}%
\providecommand \@sanitize@url [0]{\catcode `\\12\catcode `\$12\catcode
  `\&12\catcode `\#12\catcode `\^12\catcode `\_12\catcode `\%12\relax}%
\providecommand \@@startlink[1]{}%
\providecommand \@@endlink[0]{}%
\providecommand \url  [0]{\begingroup\@sanitize@url \@url }%
\providecommand \@url [1]{\endgroup\@href {#1}{\urlprefix }}%
\providecommand \urlprefix  [0]{URL }%
\providecommand \Eprint [0]{\href }%
\providecommand \doibase [0]{https://doi.org/}%
\providecommand \selectlanguage [0]{\@gobble}%
\providecommand \bibinfo  [0]{\@secondoftwo}%
\providecommand \bibfield  [0]{\@secondoftwo}%
\providecommand \translation [1]{[#1]}%
\providecommand \BibitemOpen [0]{}%
\providecommand \bibitemStop [0]{}%
\providecommand \bibitemNoStop [0]{.\EOS\space}%
\providecommand \EOS [0]{\spacefactor3000\relax}%
\providecommand \BibitemShut  [1]{\csname bibitem#1\endcsname}%
\let\auto@bib@innerbib\@empty
%</preamble>
\bibitem [{\citenamefont {McArdle}\ \emph {et~al.}(2020)\citenamefont
  {McArdle}, \citenamefont {Endo}, \citenamefont {Aspuru-Guzik}, \citenamefont
  {Benjamin},\ and\ \citenamefont {Yuan}}]{mcardle2020quantum}%
  \BibitemOpen
  \bibfield  {author} {\bibinfo {author} {\bibfnamefont {S.}~\bibnamefont
  {McArdle}}, \bibinfo {author} {\bibfnamefont {S.}~\bibnamefont {Endo}},
  \bibinfo {author} {\bibfnamefont {A.}~\bibnamefont {Aspuru-Guzik}}, \bibinfo
  {author} {\bibfnamefont {S.~C.}\ \bibnamefont {Benjamin}},\ and\ \bibinfo
  {author} {\bibfnamefont {X.}~\bibnamefont {Yuan}},\ }\href
  {https://doi.org/10.1103/RevModPhys.92.015003} {\bibfield  {journal}
  {\bibinfo  {journal} {Rev. Mod. Phys.}\ }\textbf {\bibinfo {volume} {92}},\
  \bibinfo {pages} {015003} (\bibinfo {year} {2020})}\BibitemShut {NoStop}%
\bibitem [{\citenamefont {Elfving}\ \emph {et~al.}(2020)\citenamefont
  {Elfving}, \citenamefont {Broer}, \citenamefont {Webber}, \citenamefont
  {Gavartin}, \citenamefont {Halls}, \citenamefont {Lorton},\ and\
  \citenamefont {Bochevarov}}]{elfving2020will}%
  \BibitemOpen
  \bibfield  {author} {\bibinfo {author} {\bibfnamefont {V.~E.}\ \bibnamefont
  {Elfving}}, \bibinfo {author} {\bibfnamefont {B.~W.}\ \bibnamefont {Broer}},
  \bibinfo {author} {\bibfnamefont {M.}~\bibnamefont {Webber}}, \bibinfo
  {author} {\bibfnamefont {J.}~\bibnamefont {Gavartin}}, \bibinfo {author}
  {\bibfnamefont {M.~D.}\ \bibnamefont {Halls}}, \bibinfo {author}
  {\bibfnamefont {K.~P.}\ \bibnamefont {Lorton}},\ and\ \bibinfo {author}
  {\bibfnamefont {A.}~\bibnamefont {Bochevarov}},\ }\href@noop {} {\bibfield
  {journal} {\bibinfo  {journal} {arXiv preprint arXiv:2009.12472}\ } (\bibinfo
  {year} {2020})}\BibitemShut {NoStop}%
\bibitem [{\citenamefont {Stamatopoulos}\ \emph {et~al.}(2022)\citenamefont
  {Stamatopoulos}, \citenamefont {Mazzola}, \citenamefont {Woerner},\ and\
  \citenamefont {Zeng}}]{stamatopoulos2022towards}%
  \BibitemOpen
  \bibfield  {author} {\bibinfo {author} {\bibfnamefont {N.}~\bibnamefont
  {Stamatopoulos}}, \bibinfo {author} {\bibfnamefont {G.}~\bibnamefont
  {Mazzola}}, \bibinfo {author} {\bibfnamefont {S.}~\bibnamefont {Woerner}},\
  and\ \bibinfo {author} {\bibfnamefont {W.~J.}\ \bibnamefont {Zeng}},\
  }\href@noop {} {\bibfield  {journal} {\bibinfo  {journal} {Quantum}\ }\textbf
  {\bibinfo {volume} {6}},\ \bibinfo {pages} {770} (\bibinfo {year}
  {2022})}\BibitemShut {NoStop}%
\bibitem [{\citenamefont {Blekos}\ \emph {et~al.}(2023)\citenamefont {Blekos},
  \citenamefont {Brand}, \citenamefont {Ceschini}, \citenamefont {Chou},
  \citenamefont {Li}, \citenamefont {Pandya},\ and\ \citenamefont
  {Summer}}]{blekos2023review}%
  \BibitemOpen
  \bibfield  {author} {\bibinfo {author} {\bibfnamefont {K.}~\bibnamefont
  {Blekos}}, \bibinfo {author} {\bibfnamefont {D.}~\bibnamefont {Brand}},
  \bibinfo {author} {\bibfnamefont {A.}~\bibnamefont {Ceschini}}, \bibinfo
  {author} {\bibfnamefont {C.-H.}\ \bibnamefont {Chou}}, \bibinfo {author}
  {\bibfnamefont {R.-H.}\ \bibnamefont {Li}}, \bibinfo {author} {\bibfnamefont
  {K.}~\bibnamefont {Pandya}},\ and\ \bibinfo {author} {\bibfnamefont
  {A.}~\bibnamefont {Summer}},\ }\href@noop {} {\bibfield  {journal} {\bibinfo
  {journal} {arXiv preprint arXiv:2306.09198}\ } (\bibinfo {year}
  {2023})}\BibitemShut {NoStop}%
\bibitem [{\citenamefont {Bhattacharjee}\ and\ \citenamefont
  {Chattopadhyay}(2017)}]{Bhattacharjee2017}%
  \BibitemOpen
  \bibfield  {author} {\bibinfo {author} {\bibfnamefont {D.}~\bibnamefont
  {Bhattacharjee}}\ and\ \bibinfo {author} {\bibfnamefont {A.}~\bibnamefont
  {Chattopadhyay}},\ }\href@noop {} {\bibfield  {journal} {\bibinfo  {journal}
  {arXiv preprint arXiv:1703.08540}\ } (\bibinfo {year} {2017})}\BibitemShut
  {NoStop}%
\bibitem [{\citenamefont {Siraichi}\ \emph {et~al.}(2019)\citenamefont
  {Siraichi}, \citenamefont {dos Santos}, \citenamefont {Collange},\ and\
  \citenamefont {Pereira}}]{Siraichi2019}%
  \BibitemOpen
  \bibfield  {author} {\bibinfo {author} {\bibfnamefont {M.~Y.}\ \bibnamefont
  {Siraichi}}, \bibinfo {author} {\bibfnamefont {V.~F.}\ \bibnamefont {dos
  Santos}}, \bibinfo {author} {\bibfnamefont {C.}~\bibnamefont {Collange}},\
  and\ \bibinfo {author} {\bibfnamefont {F.~M.~Q.}\ \bibnamefont {Pereira}},\
  }\href@noop {} {\bibfield  {journal} {\bibinfo  {journal} {Proceedings of the
  ACM on Programming Languages}\ }\textbf {\bibinfo {volume} {3}},\ \bibinfo
  {pages} {1 } (\bibinfo {year} {2019})}\BibitemShut {NoStop}%
\bibitem [{\citenamefont {Miltzow}\ \emph {et~al.}(2016)\citenamefont
  {Miltzow}, \citenamefont {Narins}, \citenamefont {Okamoto}, \citenamefont
  {Rote}, \citenamefont {Thomas},\ and\ \citenamefont {Uno}}]{Miltzow2016}%
  \BibitemOpen
  \bibfield  {author} {\bibinfo {author} {\bibfnamefont {T.}~\bibnamefont
  {Miltzow}}, \bibinfo {author} {\bibfnamefont {L.}~\bibnamefont {Narins}},
  \bibinfo {author} {\bibfnamefont {Y.}~\bibnamefont {Okamoto}}, \bibinfo
  {author} {\bibfnamefont {G.}~\bibnamefont {Rote}}, \bibinfo {author}
  {\bibfnamefont {A.}~\bibnamefont {Thomas}},\ and\ \bibinfo {author}
  {\bibfnamefont {T.}~\bibnamefont {Uno}},\ }in\ \href
  {https://doi.org/10.4230/LIPIcs.ESA.2016.66} {\emph {\bibinfo {booktitle}
  {24th Annual European Symposium on Algorithms (ESA 2016)}}},\ \bibinfo
  {series} {Leibniz International Proceedings in Informatics (LIPIcs)},
  Vol.~\bibinfo {volume} {57},\ \bibinfo {editor} {edited by\ \bibinfo {editor}
  {\bibfnamefont {P.}~\bibnamefont {Sankowski}}\ and\ \bibinfo {editor}
  {\bibfnamefont {C.}~\bibnamefont {Zaroliagis}}}\ (\bibinfo  {publisher}
  {Schloss Dagstuhl--Leibniz-Zentrum fuer Informatik},\ \bibinfo {address}
  {Dagstuhl, Germany},\ \bibinfo {year} {2016})\ pp.\ \bibinfo {pages}
  {66:1--66:15}\BibitemShut {NoStop}%
\bibitem [{\citenamefont {Karp}(1972)}]{Karp1972}%
  \BibitemOpen
  \bibfield  {author} {\bibinfo {author} {\bibfnamefont {R.~M.}\ \bibnamefont
  {Karp}},\ }\bibinfo {title} {Reducibility among combinatorial problems},\ in\
  \href {https://doi.org/10.1007/978-1-4684-2001-2_9} {\emph {\bibinfo
  {booktitle} {Complexity of Computer Computations: Proceedings of a symposium
  on the Complexity of Computer Computations, held March 20--22, 1972, at the
  IBM Thomas J. Watson Research Center, Yorktown Heights, New York, and
  sponsored by the Office of Naval Research, Mathematics Program, IBM World
  Trade Corporation, and the IBM Research Mathematical Sciences Department}}},\
  \bibinfo {editor} {edited by\ \bibinfo {editor} {\bibfnamefont {R.~E.}\
  \bibnamefont {Miller}}, \bibinfo {editor} {\bibfnamefont {J.~W.}\
  \bibnamefont {Thatcher}},\ and\ \bibinfo {editor} {\bibfnamefont {J.~D.}\
  \bibnamefont {Bohlinger}}}\ (\bibinfo  {publisher} {Springer US},\ \bibinfo
  {address} {Boston, MA},\ \bibinfo {year} {1972})\ pp.\ \bibinfo {pages}
  {85--103}\BibitemShut {NoStop}%
\bibitem [{\citenamefont {Murali}\ \emph {et~al.}(2019)\citenamefont {Murali},
  \citenamefont {Baker}, \citenamefont {Javadi-Abhari}, \citenamefont {Chong},\
  and\ \citenamefont {Martonosi}}]{Murali2019}%
  \BibitemOpen
  \bibfield  {author} {\bibinfo {author} {\bibfnamefont {P.}~\bibnamefont
  {Murali}}, \bibinfo {author} {\bibfnamefont {J.~M.}\ \bibnamefont {Baker}},
  \bibinfo {author} {\bibfnamefont {A.}~\bibnamefont {Javadi-Abhari}}, \bibinfo
  {author} {\bibfnamefont {F.~T.}\ \bibnamefont {Chong}},\ and\ \bibinfo
  {author} {\bibfnamefont {M.}~\bibnamefont {Martonosi}},\ }\href@noop {}
  {\bibfield  {journal} {\bibinfo  {journal} {Proceedings of the Twenty-Fourth
  International Conference on Architectural Support for Programming Languages
  and Operating Systems}\ } (\bibinfo {year} {2019})}\BibitemShut {NoStop}%
\bibitem [{\citenamefont {Tannu}\ and\ \citenamefont
  {Qureshi}(2019)}]{Tannu2019}%
  \BibitemOpen
  \bibfield  {author} {\bibinfo {author} {\bibfnamefont {S.~S.}\ \bibnamefont
  {Tannu}}\ and\ \bibinfo {author} {\bibfnamefont {M.~K.}\ \bibnamefont
  {Qureshi}},\ }in\ \href {https://doi.org/10.1145/3297858.3304007} {\emph
  {\bibinfo {booktitle} {Proceedings of the Twenty-Fourth International
  Conference on Architectural Support for Programming Languages and Operating
  Systems}}},\ \bibinfo {series and number} {ASPLOS '19}\ (\bibinfo
  {publisher} {Association for Computing Machinery},\ \bibinfo {address} {New
  York, NY, USA},\ \bibinfo {year} {2019})\ p.\ \bibinfo {pages}
  {987–999}\BibitemShut {NoStop}%
\bibitem [{\citenamefont {Ash-Saki}\ \emph {et~al.}(2019)\citenamefont
  {Ash-Saki}, \citenamefont {Alam},\ and\ \citenamefont
  {Ghosh}}]{Ash-Saki2019}%
  \BibitemOpen
  \bibfield  {author} {\bibinfo {author} {\bibfnamefont {A.}~\bibnamefont
  {Ash-Saki}}, \bibinfo {author} {\bibfnamefont {M.}~\bibnamefont {Alam}},\
  and\ \bibinfo {author} {\bibfnamefont {S.}~\bibnamefont {Ghosh}},\ }in\ \href
  {https://doi.org/10.1145/3316781.3317888} {\emph {\bibinfo {booktitle}
  {Proceedings of the 56th Annual Design Automation Conference 2019}}},\
  \bibinfo {series and number} {DAC '19}\ (\bibinfo  {publisher} {Association
  for Computing Machinery},\ \bibinfo {address} {New York, NY, USA},\ \bibinfo
  {year} {2019})\BibitemShut {NoStop}%
\bibitem [{\citenamefont {Niu}\ \emph {et~al.}(2020)\citenamefont {Niu},
  \citenamefont {Suau}, \citenamefont {Staffelbach},\ and\ \citenamefont
  {Todri-Sanial}}]{Siyuan2020}%
  \BibitemOpen
  \bibfield  {author} {\bibinfo {author} {\bibfnamefont {S.}~\bibnamefont
  {Niu}}, \bibinfo {author} {\bibfnamefont {A.}~\bibnamefont {Suau}}, \bibinfo
  {author} {\bibfnamefont {G.}~\bibnamefont {Staffelbach}},\ and\ \bibinfo
  {author} {\bibfnamefont {A.}~\bibnamefont {Todri-Sanial}},\ }\href@noop {}
  {\bibfield  {journal} {\bibinfo  {journal} {IEEE Transactions on Quantum
  Engineering}\ }\textbf {\bibinfo {volume} {1}},\ \bibinfo {pages} {1}
  (\bibinfo {year} {2020})}\BibitemShut {NoStop}%
\bibitem [{\citenamefont {Hashim}\ \emph {et~al.}(2022)\citenamefont {Hashim},
  \citenamefont {Rines}, \citenamefont {Omole}, \citenamefont {Naik},
  \citenamefont {Kreikebaum}, \citenamefont {Santiago}, \citenamefont {Chong},
  \citenamefont {Siddiqi},\ and\ \citenamefont
  {Gokhale}}]{hashim2022optimized}%
  \BibitemOpen
  \bibfield  {author} {\bibinfo {author} {\bibfnamefont {A.}~\bibnamefont
  {Hashim}}, \bibinfo {author} {\bibfnamefont {R.}~\bibnamefont {Rines}},
  \bibinfo {author} {\bibfnamefont {V.}~\bibnamefont {Omole}}, \bibinfo
  {author} {\bibfnamefont {R.~K.}\ \bibnamefont {Naik}}, \bibinfo {author}
  {\bibfnamefont {J.~M.}\ \bibnamefont {Kreikebaum}}, \bibinfo {author}
  {\bibfnamefont {D.~I.}\ \bibnamefont {Santiago}}, \bibinfo {author}
  {\bibfnamefont {F.~T.}\ \bibnamefont {Chong}}, \bibinfo {author}
  {\bibfnamefont {I.}~\bibnamefont {Siddiqi}},\ and\ \bibinfo {author}
  {\bibfnamefont {P.}~\bibnamefont {Gokhale}},\ }\href
  {https://doi.org/10.1103/PhysRevResearch.4.033028} {\bibfield  {journal}
  {\bibinfo  {journal} {Phys. Rev. Res.}\ }\textbf {\bibinfo {volume} {4}},\
  \bibinfo {pages} {033028} (\bibinfo {year} {2022})}\BibitemShut {NoStop}%
\bibitem [{\citenamefont {Acampora}\ and\ \citenamefont
  {Schiattarella}(2021)}]{Acampora2021}%
  \BibitemOpen
  \bibfield  {author} {\bibinfo {author} {\bibfnamefont {G.}~\bibnamefont
  {Acampora}}\ and\ \bibinfo {author} {\bibfnamefont {R.}~\bibnamefont
  {Schiattarella}},\ }\href@noop {} {\bibfield  {journal} {\bibinfo  {journal}
  {Neural Computing and Applications}\ }\textbf {\bibinfo {volume} {33}},\
  \bibinfo {pages} {13723 } (\bibinfo {year} {2021})}\BibitemShut {NoStop}%
\bibitem [{\citenamefont {Pozzi}\ \emph {et~al.}(2022)\citenamefont {Pozzi},
  \citenamefont {Herbert}, \citenamefont {Sengupta},\ and\ \citenamefont
  {Mullins}}]{Pozzi2022}%
  \BibitemOpen
  \bibfield  {author} {\bibinfo {author} {\bibfnamefont {M.~G.}\ \bibnamefont
  {Pozzi}}, \bibinfo {author} {\bibfnamefont {S.~J.}\ \bibnamefont {Herbert}},
  \bibinfo {author} {\bibfnamefont {A.}~\bibnamefont {Sengupta}},\ and\
  \bibinfo {author} {\bibfnamefont {R.~D.}\ \bibnamefont {Mullins}},\
  }\bibfield  {journal} {\bibinfo  {journal} {ACM Transactions on Quantum
  Computing}\ }\textbf {\bibinfo {volume} {3}},\ \href
  {https://doi.org/10.1145/3520434} {10.1145/3520434} (\bibinfo {year}
  {2022})\BibitemShut {NoStop}%
\bibitem [{\citenamefont {Hirata}\ \emph {et~al.}(2011)\citenamefont {Hirata},
  \citenamefont {Nakanishi}, \citenamefont {Yamashita},\ and\ \citenamefont
  {Nakashima}}]{hirata2011efficient}%
  \BibitemOpen
  \bibfield  {author} {\bibinfo {author} {\bibfnamefont {Y.}~\bibnamefont
  {Hirata}}, \bibinfo {author} {\bibfnamefont {M.}~\bibnamefont {Nakanishi}},
  \bibinfo {author} {\bibfnamefont {S.}~\bibnamefont {Yamashita}},\ and\
  \bibinfo {author} {\bibfnamefont {Y.}~\bibnamefont {Nakashima}},\ }\href@noop
  {} {\bibfield  {journal} {\bibinfo  {journal} {Quantum Information \&
  Computation}\ }\textbf {\bibinfo {volume} {11}},\ \bibinfo {pages} {142}
  (\bibinfo {year} {2011})}\BibitemShut {NoStop}%
\bibitem [{\citenamefont {Kivlichan}\ \emph {et~al.}(2018)\citenamefont
  {Kivlichan}, \citenamefont {McClean}, \citenamefont {Wiebe}, \citenamefont
  {Gidney}, \citenamefont {Aspuru-Guzik}, \citenamefont {Chan},\ and\
  \citenamefont {Babbush}}]{Kivlichan2018quantum}%
  \BibitemOpen
  \bibfield  {author} {\bibinfo {author} {\bibfnamefont {I.~D.}\ \bibnamefont
  {Kivlichan}}, \bibinfo {author} {\bibfnamefont {J.}~\bibnamefont {McClean}},
  \bibinfo {author} {\bibfnamefont {N.}~\bibnamefont {Wiebe}}, \bibinfo
  {author} {\bibfnamefont {C.}~\bibnamefont {Gidney}}, \bibinfo {author}
  {\bibfnamefont {A.}~\bibnamefont {Aspuru-Guzik}}, \bibinfo {author}
  {\bibfnamefont {G.~K.-L.}\ \bibnamefont {Chan}},\ and\ \bibinfo {author}
  {\bibfnamefont {R.}~\bibnamefont {Babbush}},\ }\href
  {https://doi.org/10.1103/PhysRevLett.120.110501} {\bibfield  {journal}
  {\bibinfo  {journal} {Phys. Rev. Lett.}\ }\textbf {\bibinfo {volume} {120}},\
  \bibinfo {pages} {110501} (\bibinfo {year} {2018})}\BibitemShut {NoStop}%
\bibitem [{\citenamefont {O'Gorman}\ \emph {et~al.}(2019)\citenamefont
  {O'Gorman}, \citenamefont {Huggins}, \citenamefont {Rieffel},\ and\
  \citenamefont {Whaley}}]{o2019generalized}%
  \BibitemOpen
  \bibfield  {author} {\bibinfo {author} {\bibfnamefont {B.}~\bibnamefont
  {O'Gorman}}, \bibinfo {author} {\bibfnamefont {W.~J.}\ \bibnamefont
  {Huggins}}, \bibinfo {author} {\bibfnamefont {E.~G.}\ \bibnamefont
  {Rieffel}},\ and\ \bibinfo {author} {\bibfnamefont {K.~B.}\ \bibnamefont
  {Whaley}},\ }\href@noop {} {\bibfield  {journal} {\bibinfo  {journal} {arXiv
  preprint arXiv:1905.05118}\ } (\bibinfo {year} {2019})}\BibitemShut {NoStop}%
\bibitem [{\citenamefont {Steiger}\ \emph {et~al.}(2019)\citenamefont
  {Steiger}, \citenamefont {Häner},\ and\ \citenamefont
  {Troyer}}]{Steiger2019}%
  \BibitemOpen
  \bibfield  {author} {\bibinfo {author} {\bibfnamefont {D.~S.}\ \bibnamefont
  {Steiger}}, \bibinfo {author} {\bibfnamefont {T.}~\bibnamefont {Häner}},\
  and\ \bibinfo {author} {\bibfnamefont {M.}~\bibnamefont {Troyer}},\ }\href
  {https://doi.org/https://doi.org/10.1016/j.micpro.2019.02.003} {\bibfield
  {journal} {\bibinfo  {journal} {Microprocessors and Microsystems}\ }\textbf
  {\bibinfo {volume} {66}},\ \bibinfo {pages} {81} (\bibinfo {year}
  {2019})}\BibitemShut {NoStop}%
\bibitem [{\citenamefont {Weidenfeller}\ \emph {et~al.}(2022)\citenamefont
  {Weidenfeller}, \citenamefont {Valor}, \citenamefont {Gacon}, \citenamefont
  {Tornow}, \citenamefont {Bello}, \citenamefont {Woerner},\ and\ \citenamefont
  {Egger}}]{Weidenfeller2022}%
  \BibitemOpen
  \bibfield  {author} {\bibinfo {author} {\bibfnamefont {J.}~\bibnamefont
  {Weidenfeller}}, \bibinfo {author} {\bibfnamefont {L.~C.}\ \bibnamefont
  {Valor}}, \bibinfo {author} {\bibfnamefont {J.}~\bibnamefont {Gacon}},
  \bibinfo {author} {\bibfnamefont {C.}~\bibnamefont {Tornow}}, \bibinfo
  {author} {\bibfnamefont {L.}~\bibnamefont {Bello}}, \bibinfo {author}
  {\bibfnamefont {S.}~\bibnamefont {Woerner}},\ and\ \bibinfo {author}
  {\bibfnamefont {D.~J.}\ \bibnamefont {Egger}},\ }\href
  {https://doi.org/10.22331/q-2022-12-07-870} {\bibfield  {journal} {\bibinfo
  {journal} {{Quantum}}\ }\textbf {\bibinfo {volume} {6}},\ \bibinfo {pages}
  {870} (\bibinfo {year} {2022})}\BibitemShut {NoStop}%
\bibitem [{\citenamefont {Zhu}\ \emph {et~al.}(2020)\citenamefont {Zhu},
  \citenamefont {Guan},\ and\ \citenamefont {Cheng}}]{Zhu2020}%
  \BibitemOpen
  \bibfield  {author} {\bibinfo {author} {\bibfnamefont {P.}~\bibnamefont
  {Zhu}}, \bibinfo {author} {\bibfnamefont {Z.}~\bibnamefont {Guan}},\ and\
  \bibinfo {author} {\bibfnamefont {X.}~\bibnamefont {Cheng}},\ }\href
  {https://doi.org/10.1109/TCAD.2020.2970594} {\bibfield  {journal} {\bibinfo
  {journal} {IEEE Transactions on Computer-Aided Design of Integrated Circuits
  and Systems}\ }\textbf {\bibinfo {volume} {39}},\ \bibinfo {pages} {4721}
  (\bibinfo {year} {2020})}\BibitemShut {NoStop}%
\bibitem [{\citenamefont {Cowtan}\ \emph {et~al.}(2019)\citenamefont {Cowtan},
  \citenamefont {Dilkes}, \citenamefont {Duncan}, \citenamefont {Krajenbrink},
  \citenamefont {Simmons},\ and\ \citenamefont {Sivarajah}}]{Cowtan2019}%
  \BibitemOpen
  \bibfield  {author} {\bibinfo {author} {\bibfnamefont {A.}~\bibnamefont
  {Cowtan}}, \bibinfo {author} {\bibfnamefont {S.}~\bibnamefont {Dilkes}},
  \bibinfo {author} {\bibfnamefont {R.}~\bibnamefont {Duncan}}, \bibinfo
  {author} {\bibfnamefont {A.}~\bibnamefont {Krajenbrink}}, \bibinfo {author}
  {\bibfnamefont {W.}~\bibnamefont {Simmons}},\ and\ \bibinfo {author}
  {\bibfnamefont {S.}~\bibnamefont {Sivarajah}},\ }in\ \href@noop {} {\emph
  {\bibinfo {booktitle} {Theory of Quantum Computation, Communication, and
  Cryptography}}}\ (\bibinfo {year} {2019})\BibitemShut {NoStop}%
\bibitem [{\citenamefont {Childs}\ \emph {et~al.}(2019)\citenamefont {Childs},
  \citenamefont {Schoute},\ and\ \citenamefont {Unsal}}]{Childs2019}%
  \BibitemOpen
  \bibfield  {author} {\bibinfo {author} {\bibfnamefont {A.~M.}\ \bibnamefont
  {Childs}}, \bibinfo {author} {\bibfnamefont {E.}~\bibnamefont {Schoute}},\
  and\ \bibinfo {author} {\bibfnamefont {C.~M.}\ \bibnamefont {Unsal}},\
  }\href@noop {} {\bibfield  {journal} {\bibinfo  {journal} {ArXiv}\ }\textbf
  {\bibinfo {volume} {abs/1902.09102}} (\bibinfo {year} {2019})}\BibitemShut
  {NoStop}%
\bibitem [{\citenamefont {Wille}\ \emph {et~al.}(2019)\citenamefont {Wille},
  \citenamefont {Burgholzer},\ and\ \citenamefont {Zulehner}}]{Wille2019}%
  \BibitemOpen
  \bibfield  {author} {\bibinfo {author} {\bibfnamefont {R.}~\bibnamefont
  {Wille}}, \bibinfo {author} {\bibfnamefont {L.}~\bibnamefont {Burgholzer}},\
  and\ \bibinfo {author} {\bibfnamefont {A.}~\bibnamefont {Zulehner}},\
  }\href@noop {} {\bibfield  {journal} {\bibinfo  {journal} {2019 56th ACM/IEEE
  Design Automation Conference (DAC)}\ ,\ \bibinfo {pages} {1}} (\bibinfo
  {year} {2019})}\BibitemShut {NoStop}%
\bibitem [{\citenamefont {Zulehner}\ \emph {et~al.}(2017)\citenamefont
  {Zulehner}, \citenamefont {Paler},\ and\ \citenamefont
  {Wille}}]{Zulehner2017}%
  \BibitemOpen
  \bibfield  {author} {\bibinfo {author} {\bibfnamefont {A.}~\bibnamefont
  {Zulehner}}, \bibinfo {author} {\bibfnamefont {A.}~\bibnamefont {Paler}},\
  and\ \bibinfo {author} {\bibfnamefont {R.}~\bibnamefont {Wille}},\
  }\href@noop {} {\bibfield  {journal} {\bibinfo  {journal} {IEEE Transactions
  on Computer-Aided Design of Integrated Circuits and Systems}\ }\textbf
  {\bibinfo {volume} {38}},\ \bibinfo {pages} {1226} (\bibinfo {year}
  {2017})}\BibitemShut {NoStop}%
\bibitem [{\citenamefont {Siraichi}\ \emph {et~al.}(2018)\citenamefont
  {Siraichi}, \citenamefont {dos Santos}, \citenamefont {Collange},\ and\
  \citenamefont {Pereira}}]{Siraichi2018}%
  \BibitemOpen
  \bibfield  {author} {\bibinfo {author} {\bibfnamefont {M.~Y.}\ \bibnamefont
  {Siraichi}}, \bibinfo {author} {\bibfnamefont {V.~F.}\ \bibnamefont {dos
  Santos}}, \bibinfo {author} {\bibfnamefont {C.}~\bibnamefont {Collange}},\
  and\ \bibinfo {author} {\bibfnamefont {F.~M.~Q.}\ \bibnamefont {Pereira}},\
  }\href@noop {} {\bibfield  {journal} {\bibinfo  {journal} {Proceedings of the
  2018 International Symposium on Code Generation and Optimization}\ }
  (\bibinfo {year} {2018})}\BibitemShut {NoStop}%
\bibitem [{\citenamefont {Li}\ \emph {et~al.}(2019{\natexlab{a}})\citenamefont
  {Li}, \citenamefont {Ding},\ and\ \citenamefont {Xie}}]{Li2019}%
  \BibitemOpen
  \bibfield  {author} {\bibinfo {author} {\bibfnamefont {G.}~\bibnamefont
  {Li}}, \bibinfo {author} {\bibfnamefont {Y.}~\bibnamefont {Ding}},\ and\
  \bibinfo {author} {\bibfnamefont {Y.}~\bibnamefont {Xie}},\ }in\ \href
  {https://doi.org/10.1145/3297858.3304023} {\emph {\bibinfo {booktitle}
  {Proceedings of the Twenty-Fourth International Conference on Architectural
  Support for Programming Languages and Operating Systems, {ASPLOS} 2019,
  Providence, RI, USA, April 13-17, 2019}}},\ \bibinfo {editor} {edited by\
  \bibinfo {editor} {\bibfnamefont {I.}~\bibnamefont {Bahar}}, \bibinfo
  {editor} {\bibfnamefont {M.}~\bibnamefont {Herlihy}}, \bibinfo {editor}
  {\bibfnamefont {E.}~\bibnamefont {Witchel}},\ and\ \bibinfo {editor}
  {\bibfnamefont {A.~R.}\ \bibnamefont {Lebeck}}}\ (\bibinfo  {publisher}
  {{ACM}},\ \bibinfo {year} {2019})\ pp.\ \bibinfo {pages}
  {1001--1014}\BibitemShut {NoStop}%
\bibitem [{\citenamefont {Li}\ \emph {et~al.}(2019{\natexlab{b}})\citenamefont
  {Li}, \citenamefont {Ding},\ and\ \citenamefont {Xie}}]{Gushu2019}%
  \BibitemOpen
  \bibfield  {author} {\bibinfo {author} {\bibfnamefont {G.}~\bibnamefont
  {Li}}, \bibinfo {author} {\bibfnamefont {Y.}~\bibnamefont {Ding}},\ and\
  \bibinfo {author} {\bibfnamefont {Y.}~\bibnamefont {Xie}},\ }\href@noop {}
  {\bibinfo {title} {Tackling the qubit mapping problem for nisq-era quantum
  devices}} (\bibinfo {year} {2019}{\natexlab{b}}),\ \Eprint
  {https://arxiv.org/abs/1809.02573} {arXiv:1809.02573 [cs.ET]} \BibitemShut
  {NoStop}%
\bibitem [{\citenamefont {Lao}\ and\ \citenamefont
  {Browne}(2021)}]{Lingling2021}%
  \BibitemOpen
  \bibfield  {author} {\bibinfo {author} {\bibfnamefont {L.}~\bibnamefont
  {Lao}}\ and\ \bibinfo {author} {\bibfnamefont {D.~E.}\ \bibnamefont
  {Browne}},\ }\href@noop {} {\bibinfo {title} {2qan: A quantum compiler for
  2-local qubit hamiltonian simulation algorithms}} (\bibinfo {year} {2021}),\
  \Eprint {https://arxiv.org/abs/2108.02099} {arXiv:2108.02099 [quant-ph]}
  \BibitemShut {NoStop}%
\bibitem [{\citenamefont {Farhi}\ \emph {et~al.}(2014)\citenamefont {Farhi},
  \citenamefont {Goldstone},\ and\ \citenamefont {Gutmann}}]{Farhi2014}%
  \BibitemOpen
  \bibfield  {author} {\bibinfo {author} {\bibfnamefont {E.}~\bibnamefont
  {Farhi}}, \bibinfo {author} {\bibfnamefont {J.}~\bibnamefont {Goldstone}},\
  and\ \bibinfo {author} {\bibfnamefont {S.}~\bibnamefont {Gutmann}},\
  }\href@noop {} {\bibinfo {title} {A quantum approximate optimization
  algorithm}} (\bibinfo {year} {2014}),\ \Eprint
  {https://arxiv.org/abs/1411.4028} {arXiv:1411.4028 [quant-ph]} \BibitemShut
  {NoStop}%
\bibitem [{\citenamefont {Vizing}(1965)}]{Vizing1965}%
  \BibitemOpen
  \bibfield  {author} {\bibinfo {author} {\bibfnamefont {V.~G.}\ \bibnamefont
  {Vizing}},\ }\href {https://doi.org/10.1007/bf01885700} {\bibfield  {journal}
  {\bibinfo  {journal} {Cybernetics}\ }\textbf {\bibinfo {volume} {1}},\
  \bibinfo {pages} {32} (\bibinfo {year} {1965})}\BibitemShut {NoStop}%
\bibitem [{\citenamefont {Lao}\ and\ \citenamefont {Browne}(2022)}]{Lao2022}%
  \BibitemOpen
  \bibfield  {author} {\bibinfo {author} {\bibfnamefont {L.}~\bibnamefont
  {Lao}}\ and\ \bibinfo {author} {\bibfnamefont {D.~E.}\ \bibnamefont
  {Browne}},\ }in\ \href {https://doi.org/10.1145/3470496.3527394} {\emph
  {\bibinfo {booktitle} {Proceedings of the 49th Annual International Symposium
  on Computer Architecture}}},\ \bibinfo {series and number} {ISCA '22}\
  (\bibinfo  {publisher} {Association for Computing Machinery},\ \bibinfo
  {address} {New York, NY, USA},\ \bibinfo {year} {2022})\ p.\ \bibinfo {pages}
  {351–365}\BibitemShut {NoStop}%
\bibitem [{\citenamefont {Bernshteyn}\ and\ \citenamefont
  {Dhawan}(2023)}]{bernshteyn2023fast}%
  \BibitemOpen
  \bibfield  {author} {\bibinfo {author} {\bibfnamefont {A.}~\bibnamefont
  {Bernshteyn}}\ and\ \bibinfo {author} {\bibfnamefont {A.}~\bibnamefont
  {Dhawan}},\ }\href@noop {} {\bibfield  {journal} {\bibinfo  {journal} {arXiv
  preprint arXiv:2303.05408}\ } (\bibinfo {year} {2023})}\BibitemShut {NoStop}%
\bibitem [{\citenamefont {Itoko}\ \emph {et~al.}(2020)\citenamefont {Itoko},
  \citenamefont {Raymond}, \citenamefont {Imamichi},\ and\ \citenamefont
  {Matsuo}}]{itoko2020optimization}%
  \BibitemOpen
  \bibfield  {author} {\bibinfo {author} {\bibfnamefont {T.}~\bibnamefont
  {Itoko}}, \bibinfo {author} {\bibfnamefont {R.}~\bibnamefont {Raymond}},
  \bibinfo {author} {\bibfnamefont {T.}~\bibnamefont {Imamichi}},\ and\
  \bibinfo {author} {\bibfnamefont {A.}~\bibnamefont {Matsuo}},\ }\href@noop {}
  {\bibfield  {journal} {\bibinfo  {journal} {Integration}\ }\textbf {\bibinfo
  {volume} {70}},\ \bibinfo {pages} {43} (\bibinfo {year} {2020})}\BibitemShut
  {NoStop}%
\end{thebibliography}%
\bibliographystyle{apsrev4-2}

\end{document}